\newif\iflong
\newif\ifshort
\newif\ifaltermain

\iflong 
\else
\shorttrue
\fi

\documentclass[letterpaper]{article} %
\usepackage{aaai24}  %
\usepackage{times}  %
\usepackage{helvet}  %
\usepackage{courier}  %
\usepackage[hyphens]{url}  %
\usepackage{natbib}  %
\usepackage{caption} %
\usepackage[noend,ruled,linesnumbered]{algorithm2e}

\SetAlFnt{\small}
\SetAlCapFnt{\small}
\SetAlCapNameFnt{\small} 
\SetAlCapHSkip{0pt}

\SetKwInput{KwI}{Input}
\SetKwInput{KwO}{Output}
\SetKw{KwR}{return}
\SetKw{KwNo}{no}
\SetKw{KwValid}{valid}
\SetKw{KwInvalid}{invalid}

\usepackage{hyperref} 
\hypersetup{%
 backref=true, 
 pagebackref=true, 
 hypertexnames=true,
 colorlinks=true,citecolor=green!35!black,linkcolor=red!60!black%
}

\SetCommentSty{mycommfont}
\SetKwComment{Comment}{$\triangleright$\ }{}	

\nocopyright %

\newcommand{\mytitle}{Algorithms and Complexity for Congested Assignments}

\usepackage{amsmath,amssymb,latexsym,amsthm}

\title{\mytitle}
\author {
    Jiehua Chen\textsuperscript{\rm 1},
    Jiong Guo\textsuperscript{\rm 2},
    Yinghui Wen\textsuperscript{\rm 2}
}
\affiliations {
    \textsuperscript{\rm 1}TU Wien, Austria\\
    \textsuperscript{\rm 2}Shangdong University, China\\
    jiehua.chen@ac.tuwien.ac.at, jguo@sdu.edu.cn, yhwen@mail.sdu.edu.cn\\
}

\usepackage{paralist}
\usepackage{color}
\usepackage[colorinlistoftodos,bordercolor=orange,backgroundcolor=orange!20,linecolor=orange,textsize=scriptsize,disable]{todonotes}

\usepackage{soul}
\usepackage{mathtools}

\newtheorem{theorem}{Theorem}
\newtheorem{definition}{Definition}
\newtheorem*{definition*}{Definition}
\newtheorem{lemma}{Lemma}

\newtheorem{claim}{Claim}[theorem]

\newtheorem{observation}{Observation}
\newtheorem{example}{Example}

\newtheorem{construction}{Construction}

\usepackage{cleveref}

\crefname{table}{Table}{Tables}
\crefname{figure}{Figure}{Figures}
\crefname{theorem}{Theorem}{Theorems}
\crefname{definition}{Definition}{Definitions}
\crefname{corollary}{Corollary}{Corollaries}
\crefname{observation}{Observation}{Observations}
\crefname{lemma}{Lemma}{Lemmas}
\crefname{example}{Example}{Examples}
\crefname{reduction}{Reduction}{Reductions}
\crefname{construction}{Construction}{Constructions}
\crefname{subsection}{Subsection}{Subsections}
\crefname{section}{Section}{Sections}
\crefname{proposition}{Proposition}{Propositions}
\crefname{algorithm}{Algorithm}{Algorithms}
\crefname{drule}{Rule}{Rules}
\crefname{claim}{Claim}{Claims}
\crefname{appendix}{Appendix}{Appendix}
\crefname{conjecture}{Conjecture}{Conjectures}

\usepackage{tikz}
\usetikzlibrary{decorations,arrows,petri,topaths,backgrounds,shapes,positioning,fit,calc,decorations.pathreplacing,patterns,intersections,decorations.pathmorphing,matrix}

\tikzset{linemarkr/.style =   {%
    opacity=.4,
    line width= 7pt,
    draw=Wcolor}} 
\tikzset{linemarkg/.style =   {decorate, decoration={snake,amplitude=.55mm,segment length=3mm}, line width= 2.5pt, draw=Ucolor}}
\tikzset{linemarkb/.style =   {line cap=round, opacity=.15, line width= 7pt, blue}}
\tikzset{linemarky/.style =   {line cap=round, opacity=.2, line width= 7pt, yellow}}

\tikzset{smalllinemarkstyle/.style={#1,opacity=.2, line width = 7pt}}
\tikzset{bpstyle/.style={#1,opacity=.5, line width = 4pt}}
\tikzset{agentnode/.style={draw, circle, fill=white, minimum size=1ex, inner sep=0pt, fill=black}}
\tikzset{labelnode/.style={inner sep=1.5pt, font=\small}}
\tikzset{cutsetcircle/.style={draw, gray, opacity=0.8, line width=1.3pt, dashed}}
\tikzset{extranode/.style={draw=blue!80!black, circle, fill=white, minimum size=1ex, inner sep=0pt, fill=blue!70!black}}

\usepackage{etoolbox} %
\newcommand{\tocommentout}[1]{%
}

\usepackage{xspace}
\newcommand{\myemph}[1]{{\color{green!40!black}\emph{#1}}}
\usepackage{bibentry}

\newcommand\MaxCong[2]{\lambda({#1,#2})}
\newcommand{\NETV}{\hat{V}}
\newcommand{\NETA}{\hat{A}}
\newcommand{\hata}{\hat{a}}
\newcommand{\hatv}{\hat{v}}
\newcommand{\specialv}{\hat{v^*}}

\newcommand{\capa}{\mathsf{c}}

\newcommand{\ExactCover}{\textsc{Exact Cover by 3-Sets}\xspace}
\newcommand{\Congested}{\textsc{Congested Assignment}\xspace}

\newcommand{\decprob}[3]{
  \begin{center}
    \begin{minipage}{0.92\linewidth}
    \begin{description}%
      \item[\textsc{#1}]
      \item[Input:]  #2
      \item[Question:] #3
    \end{description}
  \end{minipage}
  \end{center}%
}

\begin{document}

\maketitle

\begin{abstract}
  We study the congested assignment problem as introduced by \citet{BogoMoulin23}.
  We show that deciding whether a competitive assignment exists can be done in polynomial time, while deciding whether an envy-free assignment exists is NP-complete.
\end{abstract}

\section{Introduction}\label{sec:intro}

In congested assignment, we need to assign agents to posts where each agent has weak preferences over not only the posts but also together with congestions; a congestion for a post is the number of agents that are assigned to it.
These preferences are negative over congestions meaning that for an acceptable post, an agent always prefers smaller congestions. %
\citet{BogoMoulin23} introduce the concept of \myemph{competitiveness} which requires that no agent \myemph{envies} another agent (i.e., would like to take over another agent's assignment) and no agent would like to go to an empty post.
They show that competitive assignments, however, do not always exist and leave open the complexity of determining the existence of competitive assignments.
In this short paper, we show how to determine competitiveness in polynomial time.
The crucial idea is to reduce to finding a competitive assignment where no post is empty, in which case it is equivalent to envy-free assignments with no empty posts.
We show that determining the existence of envy-free assignments without any empty posts
reduces to determining whether the maximum flow of an auxiliary flow network has some desired value.
The latter problem is known to be polynomial-time solvable.
Hence, the approach can be done in polynomial time; see \cref{Alg:CP_Ties}.

One may ask whether it is also polynomial-time solvable to determine any envy-free assignment in case some posts need to be empty.
We answer this question in the negative by showing that it is NP-complete to determine the existence of any envy-free assignment (allowing empty posts); see \cref{thm:EF-NP-hard}.
We provide a polynomial reduction from the NP-complete \textsc{Exact Cover by 3 Sets} problem~\cite{GJ79}.
This means that allowing empty posts makes the problem harder.

\paragraph{Paper outline}
In \cref{sec:prelim}, we introduce necessary definitions and concepts for the paper.
In \cref{sec:CP}, we present our algorithm for competitive assignments,
while in \cref{sec:EF}, we show NP-hardness for envy-free assignments.

\section{Preliminaries}\label{sec:prelim}
Given a non-negative integer~$z$, let \myemph{$[z]$} denote the set~$\{1,\ldots,z\}$.
Let $A=\{a_1,\ldots,a_m\}$ denote a finite set of $m$ \myemph{posts}.
The input of \textsc{Congested Assignment} consists of the set~$A$, a finite set~\myemph{$V$} of $n$ agents, where each agent~$v\in V$ has a \myemph{preference list}~$\succeq_v$ (i.e., a weak order, which is transitive and complete) on the set of pairs~$A\times [n]$.
The weak order~$\succeq_v$ specifies the preferences of agent~$v$ over the posts and their congestions (i.e., the number of agents that will simultaneously occupy a post).
The agents may be indifferent between different posts, but strictly decreasing in the congestions when the same post is considered.
For instance, agent~$v$ may have $(a_1, 2)\succeq_i (a_2, 3)$, meaning that he weakly prefers assigned to post~$a_1$ together with \emph{one} other agent to post~$a_2$ together with \emph{two} other agents.
This also implies that he has $(a_1,1)\succ_v (a_2,3)$ (because $(a_1,1)\succ_v (a_1, 2)$), meaning that he strictly prefers being assigned to post~$a_1$ alone over to post~$a_2$ with two agents.
We use \myemph{$\sim_v$} to denote the symmetric part of $\succeq_v$ and \myemph{$\succ_v$} the asymmetric part.

An \myemph{assignment} of agents~$V$ to posts~$A$ is a partition~\myemph{$\Pi=(S_a)_{a\in A}$} of $V$ where $S_a$ is the set of agents assigned to post~$a$ so that every two sets~$S_a$ and $S_b$ are mutually disjoint and $\cup_{a\in A}S_a = V$. The cardinality~$|S_a|$ of $S_a$ is called the \myemph{congestion} of post~$a$.
We say that a post~$a$ is \myemph{empty} if $|S_a|=0$.
For brevity's sake, we use \myemph{$\Pi(v)$} to refer to the post that agent~$v$ is assigned to and \myemph{$\Pi(a)$} to refer to the set~$S_{a}$ of agents that are assigned to post~$a$.
We call \myemph{$\mathbf{s} = (|S_a|)_{a\in A}$} the \myemph{congestion profile} of partition~$\Pi$.

As by \citet{BogoMoulin23}, we assume that every agent's preference list has length $|V|$.
This is because competitive assignments (see below) always ensure that every agent is assigned to a post with congestion that is one of the first $|V|$ tuples in his preference list.
Accordingly, given an agent~$v$ and a post~$a$, we use \myemph{$\MaxCong{v}{a}$} to refer to the \myemph{maximum congestion} of~$v$ for post~$a$ in his preference list~$\succeq_v$.
Formally, \myemph{$\MaxCong{v}{a}=\max_{d}\{d\mid (a,d)\in \succeq_v\}$}.
By assumption, we have that $\sum_{a\in A} \MaxCong{v}{a} = |V|$ holds for every agent~$v\in V$.

\begin{definition}[Nash stable, envy-free, and competitive assignments]\label{Def:NS+EF+CP}
  Let $\Pi=(S_a)_{a\in A}$ denote an assignment for an instance~$(A, V, (\succeq_{v})_{v \in V})$ of \textsc{Congested Assignment}.
  
  We say that $\Pi$ is \myemph{Nash stable} (in short, \myemph{NS}) if no agent wishes to deviates to another post.
  Formally, $\Pi=(S_a)_{a\in A}$ is \myemph{NS} if for every agent~$v \in V$ and every post~$a\in A$ it holds that $(a^*, |S_{a^*}|) \succeq_v (a, |S_a|+1)$, where $a^*$ denotes the post that agent~$v$ is assigned to.

  We say that agent~$v$ \myemph{has an envy towards} (or simply \myemph{envies}) agent~$v_j$ if
  $(b, |S_b|) \succ_{v} (a, |S_a|)$ where $v$ is assigned to $a$ and $v_j$ to $b$.
  Accordingly, we say that $\Pi$ is \myemph{envy-free} (in short, \myemph{EF}) if no agent envies any other agent. 
  We say that $\Pi$ is \myemph{competitive} (in short, \myemph{CP}) if it is envy-free and no agent wishes to move to an empty post.
  Formally, $\Pi$ is \myemph{CP} if for every agent~$v\in V$ and every post~$a\in A$ it holds that $(a^*, |S_{a^*}|) \succeq_{v} (a, \max(|S_a|,1))$, where $a^*$ denotes the post that agent~$v$ is assigned to.
\end{definition}

The difference between NS and CP lies in whether to count the agent as one additional congestion. In NS, we count the actor, while in CP we do not (unless the considered post is empty).
By definition, CP assignments are EF, but the converse does not necessarily hold.
EF and CP coincide if no post is empty.

\begin{example}\label{ex:NS+CP}
  Consider an instance of three posts~$A=\{a_1,a_2\}$ and three agents~$V=\{v_1,v_2,v_3\}$.
  The preferences of the agents are as follows:
  \begin{align*}
    v_1\colon & (a_1, 1) \succ (a_2, 1) \sim (a_1,2),\\
    v_2\colon & (a_1,1) \sim (a_2, 1)\succ (a_1,2),\\
    v_3\colon & (a_1,1) \succ (a_1, 2)\succ (a_2,1).
  \end{align*}
  
  The maximum congestions are depicted in the following table.
  \begin{center}
    \begin{tabular}{@{}|l|l|l|}
      \hline
      $\MaxCong{v_i}{a_j}$  & $a_1$ & $a_2$\\ \hline
      $v_1, v_2, v_3$ & 2 & 1 \\\hline
    \end{tabular}
  \end{center}
  There are several NS assignments.
  For instance, $\Pi_1$ with $\Pi_1(a_1)=\{v_2,v_3\}$ and $\Pi_1(a_2)=\{v_1\}$,
  and $\Pi_2$ with $\Pi_2(a_1)=\{v_1,v_3\}$ and $\Pi_2(a_2)=\{v_2\}$.
  $\Pi_1$ is not CP (since $v_2$ prefers $(a_2,1)$ to $(a_1,2)$), but $\Pi_2$ is.
  The same holds for EF since no post is empty.
\end{example}
\citet{MilchtaichNash1996} shows that an NA assignment always exists and
finding one can be found in polynomial time by starting with an arbitrary assignment and letting agents iteratively deviate to a more preferred situation.
CP assignments, however, do not always exist as the following example shows.

\begin{example}\label{ex:noCP}
  Consider an instance of three posts~$A=\{a_1,a_2\}$ and two agents~$V=\{v_1,v_2\}$.
  The preferences of the two agents are the same: 
  \begin{align*}
    v_1, v_2\colon & (a_1, 1) \succ (a_2, 1) \succ (a_2,2).
  \end{align*}
  By the preferences over the congestions, every CP assignment must assign exactly one agent to the first post~$a_1$.
  However, the other agent will be envious.
  Hence, no CP assignments exist, but assigning both $v_1$ and $v_2$ to $a_2$ gives an EF assignment.
\end{example}

\section{Algorithms for Competitive Assignments}\label{sec:CP}

In this section, we detail our approach of deciding the existence of competitive assignments.
The main idea is to reduce to deciding the existence of a CP assignment (for an extended instance) where no post is empty.
To decide the existence of CP assignments with no posts being empty,
we construct an auxiliary network (i.e., a directed graph with arc capacities)
and iteratively increase the capacities of some arcs until we either find a flow which yields a CP assignment or decide that there is no CP assignments.
A pseudo-code description of the whole algorithm of deciding CP assignments is given in \cref{Alg:CP_Ties}, while a pseudo-code description of deciding CP assignments where no post is empty is given in \cref{Alg:CP_Ties_NoEmpty}.
\begin{algorithm}[t!]
  \caption{Determining the existence of CP assignments}
  \label{Alg:CP_Ties}
  \KwI{A instance~$I=(A, V, (\succ_v)_{v\in V})$ of \textsc{Congested Assignment}.}
  
  \KwO{A CP assignment if it exists; otherwise no.}
  
  \ForEach{$k=\max(0, m-n)$ to $m-1$\label{alg1-for}}{
    Let $I_k$ be an instance by Construction~\ref{Cons:instance} on input $(I,k)$
    
    \lIf{Algorithm~\ref{Alg:CP_Ties_NoEmpty} on input $I_k$ returns an assignment~$\Pi$}{%
      \KwR $\Pi$
    }  
  }
  \KwR \KwNo
\end{algorithm}

The main body of Algorithm~\ref{Alg:CP_Ties} is a \textbf{for}-loop, which enumerates the numbers of empty posts in a competitive assignment (if it exists).
For each number~$k$, there are two sub-steps.
First, construct an extended instance from the original one with $k$ dummy agents by using Construction~\ref{Cons:instance} (see \cref{subsec:nonemptypost}).
Second, use Algorithm~\ref{Alg:CP_Ties_NoEmpty} to solve the new instance.

In the following, we discuss the first step in \cref{subsec:nonemptypost} and the second step in \cref{subsec:noempty}, and the overall correctness in \cref{subsec:alg1}.

\subsection{Reduce to Finding CP Assignments with No Empty Posts}\label{subsec:nonemptypost}

In this section, we discuss how to reduce the problem of deciding any CP assignment to the problem deciding one where no post is empty.
The basic idea is to guess the number of empty posts (assuming the existence of CP assignments) and augment the instance with enough dummy agents and posts where any CP assignment must assign each previously empty post a distinct dummy agent. 
The core of the reduction is described in \cref{Cons:instance} below.

\begin{construction}[Extended instance]\label{Cons:instance}
  Given an instance~$I=(A, V, (\succeq_v)_{v\in V})$ of \textsc{Congested Assignment} and a number $k$ with $\max(0, n-m) \le k \le m-1$
  we construct a new instance $(A^*, V^*, (\succeq^*_{v})_{v\in V^*})$ as follows.
  Create $k+2$ dummy agents $u_1, \cdots, u_k, p_1,p_2$ and $2$ dummy posts $b_1,b_2$.
  Then, set $V^* = V \cup \{u_1, \cdots, u_k, p_1,p_2\}$ and $A^* = A \cup \{b_1, b_2\}$.
  Finally, set the preferences of the agents as follows.
  Here, $n$ and $m$ denote the number of agents and posts in the original instance, respectively,
  $v$ denotes an original agent from $I$ and $\succeq_v$ his original preference list, 
  $z\in [k]$, and $i\in [n]$.
  \begin{align*}
    u_z  \colon &(a_1,1) \sim (a_2,1) \sim \cdots \sim (a_m,1) \succ (b_1,1) \succ (b_1,2) \\
    & \succ \cdots \succ (b_1,k+n-m+2);\\
  {p_1}\colon& (b_1,1)  \succ (b_2,1) \succ (b_2,2) \succ\cdots \succ (b_2,k+n+1); \\
  {p_2}\colon& (b_2,1)  \succ (b_1,1) \succ (b_1,2) \succ\cdots \succ (b_1,k+n+1);\\
  v\colon& (\succeq_v) \succ (b_1,1) \succ  (b_1,2) \succ\cdots \succ (b_1,k+2).
\end{align*}
The following table illustrates the maximum congestion of each agent towards each post, where $a_j$ and $v$ denote the original post and agent, respectively:

\noindent
\resizebox{\linewidth}{!}{
\begin{tabular}{@{}|@{\;}c@{\;}|c@{\;}ccc@{\;}|@{\;}c@{\;}|@{\;}c@{\;}|@{}}
  \hline
    & {$a_1$} & {$a_2$} & {$\cdots$} & $a_m$ & $b_1$     & $b_2$ \\ \hline
  $u_z$ & {$1$}   & {$1$}   & {$\cdots$} & $1$   & $k+n+2-m$ & $0$   \\ \hline
  $p_1$ & {$0$} & {$0$} & {$\cdots$} & $0$ & $1$     & $k+n+1$ \\ \hline
  $p_2$ & {$0$} & {$0$} & {$\cdots$} & $0$ & $k+n+1$ & $1$     \\ \hline
  $v$ & {$\MaxCong{v}{a_1}$} & {$\MaxCong{v}{a_2}$} & {$\cdots$} & {$\MaxCong{v}{a_m}$}                                                        & $k+2$   & $0$     \\ \hline
  \end{tabular}
}\\

\hfill (of \cref{Cons:instance}) $\diamond$
\end{construction}
 
To show that the reduction is correct, we first make some observations about the CP assignments of the instance created under \cref{Cons:instance}.

\begin{observation}\label{Obs:I_k}  
  Let $I_k=(A^*, V^*, (\succeq^*_v)_{v\in V^*})$ denote the instance created by \cref{Cons:instance} with $A^*=A\cup \{b_1, b_2\}$ and $V^*=V\cup \{u_i\mid i\in [k]\}\cup \{p_1,p_2\}$. %
  Every CP assignment of $I_k$ (if it exists) satisfies the following.
  \begin{compactenum}[(1)]
    \item\label{p1} Agent~$p_1$ is assigned to $b_1$ alone.
    \item\label{p2} Agent~$p_2$ is assigned to $b_2$ alone.
    \item\label{uz} Every dummy $u_z$ with $1 \leq z \leq k$ is assigned to some $a_j \in A$ alone.
    \item\label{v} Every original agent $v_i \in V$ is assigned to some original acceptable post.
  \end{compactenum}

\end{observation}
\begin{proof}
  Let  $\Pi$ be a competitive assignment of $I_k$ with $\Pi=(S_{a})_{a\in A^*}$.
  To show statement~\eqref{p1}, we first observe that  if $p_1$ is assigned to~$b_1$, then $|S_{b_1}|=1$.
  Thus, it suffices to show that $p_1$ is assigned to~$b_1$.
  Suppose, for the sake of contradiction, that $p_1$ is assigned to~$b_2$ instead.
  Then, by the congestion of $p_2$ towards $b_2$, agent $p_2$ cannot be assigned to $b_2$ anymore.
  Since no one else considers $b_2$ acceptable, we have $S_{b_2}=\{p_1\}$, and $p_2$ envies $p_1$, a contradiction to the competitiveness.
  
  Statement~\eqref{p2} follows directly from statement~\eqref{p1} since $b_2$ is the only acceptable post left for $p_2$ and he has congestion one towards $b_2$.

  By statements~\eqref{p1}--\eqref{p2} and the maximum congestions, every dummy agent~$u_z$ can only be assigned to some original post alone, proving statement~\eqref{uz}.

  The last statement follows directly from the first three statements.
\end{proof}

Now, we show the correctness of the construction.
\begin{lemma}\label{lem:I<=>Ik}
  An instance~$I$ admits a CP assignment if and only if there exists a number~$k$, $\max(n-m, 0) \le k \le m-1$, such that the instance~$I_k$ created according to \cref{Cons:instance} admits a CP assignment where no post is empty.
\end{lemma}

\begin{proof}
  Let $I=(A,V, (\succeq_v)_{v\in V})$.
  The ``only if'' part is straightforward: Let $\Pi=(S_a)_{a\in A}$ denote a CP assignment of $I$.
  Then, let $A'$ denote the set of empty posts under $\Pi$ and set $k=|A'|$.
  Clearly, $0 \le k \le m$, and $k\ge n-m$ holds as well if $n>m$. 
  We claim that the instance~$I_k$ created according to  \cref{Cons:instance} admits a CP assignment where no post is empty.
  We construct an assignment~$\Pi_k$ for $I_k$ as follows.
  \begin{compactitem}[--]
    \item For each empty post~$a\in A'$, take a unique dummy agent~$u_z$ and assign $\Pi_k(a)=\{u_z\}$; note that there are exactly $k=|A'|$ many dummy agents.
    \item For each non-empty post~$a\in A\setminus A'$, let $\Pi_k(a)=\Pi(a)$.
    \item Let $\Pi(b_1)=\{p_1\}$ and $\Pi(b_2)=\{p_2\}$.
  \end{compactitem}
  Clearly, no post is assigned zero agents under~$\Pi_k$.
  Since no post is empty, to show competitiveness reduces to show that no agent envies another agent.
  This is clearly the case for all dummy agents and the two auxiliary agents~$p_1$ and $p_2$ since they are assigned to one of their most preferred posts alone.
  No original agent envies any other original agent or dummy agent since $\Pi$ is CP.
  No original agent envies $p_1$ and $p_2$ by their preferences extensions.
  This shows that $\Pi_k$ is CP for $I_k$, as desired.

  For the ``if'' part, let $k$ be a number between $\max(n-m, 0)$ and $m-1$ such that the created instance~$I_k$ admits a CP assignment~$\Pi_k$.
  We claim that the assignment~$\Pi$ derived from~$\Pi_k$ by restricting to the original agents from $I$ is CP for $I$.
  Formally, for each $a_j\in A$ without dummy agents, i.e., $u_z\notin \Pi_k(a_j)$,
  let $\Pi(a_j)=\Pi_k(a_j)$; otherwise let $\Pi(a_j)=\emptyset$.
  We first show that $\Pi$ is a valid assignment for $I$.
  By \cref{Obs:I_k}\eqref{uz}, every dummy agent is assigned to some original post alone.
  This means that $\Pi(a_j)\subseteq V$.
  By \cref{Obs:I_k}\eqref{v}, every original agent is assigned to an acceptable post, confirming that $\Pi$ is indeed a valid assignment for~$I$.

  Next, suppose, for the sake of contradiction, that $\Pi$ is not competitive and let $v$ and $a$ be an agent and a post such that $(a, \max(|\Pi(a)|, 1)) \succ_v (a', |\Pi(a')|)$ where $a'$ is the post that $v$ is assigned to.
  We infer that $\Pi(a)$ cannot be empty since otherwise by our construction and by \cref{Obs:I_k}\eqref{uz} we have that $\Pi_k(a)=\{u_z\}$ for some dummy agent~$u_z$.
  This further implies that $v$ envies~$u_z$ in $I_k$, a contradiction to competitiveness of $\Pi_k$.
  Since $\Pi(a)$ is not empty and $v\in \Pi(a')$, again by our construction and by \cref{Obs:I_k}\eqref{uz},
  we have that $\Pi(a)=\Pi_k(a)$ and $\Pi(a')=\Pi_k(a')$.
  By the preference extension, we obtain that $(a, |\Pi_k(a)|) \succ^*_v (a', |\Pi_k(a')|)$ a contradiction to the competitiveness of $\Pi_k$.
\end{proof}

\subsection{Determining CP Assignments with No Empty Posts}\label{subsec:noempty}

In this section, we show how to determine the existence of CP assignments with no empty posts.
The crucial idea is to reduce to finding a desired maximum flow for an auxiliary flow network.
Before we go into the details, let us describe how to construct an auxiliary network (i.e., a directed graph containing two distinguished nodes~$s$ and $t$ as source and sink, and arc capacities). 

\begin{construction}[Flow network]\label{Cons:Graph}
  Given an instance~$I=(A, V, (\succeq_v)_{v\in V})$ and a table~$T\in [|V|]^{|A|}$,
  we construct a network~$N=(G, c)$ with directed graph~$G=(\NETA\cup \NETV\cup \{s,t\}, E)$ and capacity function~$\capa\colon E(G)\to [|V|]$ as follows. %
\begin{enumerate}[(1)]
    \item For each post~$a\in A$, create a vertex $\hata$; define set~$\NETA=\{\hata\mid a\in A\}$. %
    \item For each agent~$v\in V$, create a vertex~$\hatv$; define set~$\NETV=\{\hatv\mid v\in V\}$.
    \item Create a source~$s$ and a sink~$t$.
    \item For each post~$a\in A$, create an arc~$(s, \hata)$ and set the capacity to $\capa(s, \hata)=T[a]$.
    \item For each agent~$v\in V$, create an arc~$(\hatv, t)$ and set the capacity to~$\capa(\hatv, t)=1$.
    \item For each agent~$v\in V$ and each post~$a$ with $(a, d)$ being one of the most preferred \emph{valid} tuple in $v$'s preference list, create an arc~$(\hata, \hatv)$ and set the capacity to~$\capa(\hata, \hatv)=1$; the valid status is maintained in \cref{Line:Iteration-End}. 

    \label{Item:Con2MostPreferred}
\end{enumerate}
The capacity function $\capa$ is summarized as follows.
\begin{equation*}
\capa(e)=\left\{
    \begin{aligned}
    &T[a],& \quad &\text{if } e=(s,\hata) \text{ with } a \in A,\\
    &1,& \quad &\textrm{otherwise.}\\
    \end{aligned}
\right.
\end{equation*}
\end{construction}
We need the following concepts for a network flow.

\begin{definition}[Perfect flows and the derived assignment]\label{def:flow-perfect-asignment}
  Let $N=(G, \capa)$ denote the network constructed according to \cref{Cons:Graph} for an instance~$I=(A, V, (\succeq_v)_{v\in V})$ and a table~$T\in [|V|]^{|A|}$.
  Recall that a \myemph{flow} of $N$ is a function~$f\colon E(G) \to [|V|]\cup \{0\}$ that assigns to each arc a value that does not exceed the capacity bound and overall preserves the conservation constraints: $\sum_{(x,y)\in E}f(x,y) = \sum_{(z,x)\in E}f(z,x)$ for all $x\in \NETA\cup \NETV$.
  The \myemph{value} of a flow equals the net flow into the sink~$t$. 
  We say that a flow is \myemph{perfect} if the value of $f$ is $|V|$; in other words, every arc~$(\hatv, t)$ to the sink is saturated.

  Given a perfect flow~$f$, we derive a \myemph{congested assignment~$\Pi$} for the original instance~$I$ as follows.
  For each post~$a\in A$, let $\Pi(a)=\{v\mid f(\hata, \hatv) = 1\}$.
\end{definition}

\begin{algorithm}[t]
    \caption{Determining the existence of CP assignments with no empty post}
    \label{Alg:CP_Ties_NoEmpty}
    \KwI{An instance~$I=(A, V, (\succeq_v)_{v\in V})$ of \textsc{Congested Assignment}.}
    \KwO{A CP assignment with no empty post if it exists; otherwise no.}

    $T[a] \gets 1$ for all $a \in A$ \label{Line:InitTj}
    
    $(a,d) \gets $ \KwValid for all $a \in A$ and $1\leq d\leq |V|$

    \While{$\sum_{a \in A}T[a] \leq |V|$\label{Line:SumTj<=n}}{
      \Comment{\textcolor{red}{ Phase $1$}: Deciding existence of a perfect flow}

      Let $(G=(\NETA \cup \NETV \cup \{s,t\}),E), \capa)$ denote the network constructed by \cref{Cons:Graph} on input $(I, T)$  \label{Line:phase1-start}\label{Line:Iteration-Start}

      Compute a max flow~$f$ of $(G,c)$

      \If{$f$ has value~$|V|$}{
        \KwR assignment derived from~$f$; see \cref{def:flow-perfect-asignment} \label{Line:assignment}%
      }\label{Line:phase1-end}

      \Comment{\textcolor{red}{ Phase $2$}: find an {obstruction}}

      Find a vertex~$\specialv \in \NETV$ with $f(\specialv, t)=0$\label{Line:Alg_Obs_Start}
      
      $V'\gets \{\specialv\}$ \label{Line:V'-init}

      $A' \gets \emptyset$\label{Line:A'-init}
      
      \Repeat{no $\hata$ exists}{\label{Line:Alg_Obs_While}%
        
        $\hata\gets$ a vertex in $\NETA\setminus A' \text{ w. }
         (\hata, \hatv) \in E$ for some $\hat{v}\in V'$\label{Line:A'}

        $A'\gets A'\cup \{\hata\}$

        $V'\gets V' \cup \{\hatv \in \NETV \mid f(\hata, \hatv) = 1\}$\label{Line:V'}

      }\label{Line:Alg_Obs_End}
      
     \Comment{\textcolor{red}{Phase $3$}: update invalid tuples and $T$}
     
     \ForEach{$\hat{a}\in A'$}{\label{Line:ForeachToInvalid}
       $(a, T[{a}])\gets \KwInvalid$\label{Line:SetToInvalid}
       
       $T[{a}] \gets T[{a}] + 1$ \label{Line:SetToIncrement}
     }\label{Line:Iteration-End}
   }
   \KwR \KwNo
 \end{algorithm}
 
Now, we proceed with the whole approach depicted in \cref{Alg:CP_Ties_NoEmpty}.
Briefly put, we maintain an integer table~$T$ which stores the minimum congestion for each post and
iteratively construct a flow network with capacities indicated by~$T$ and determine whether there exists a perfect flow in three phases.
We do this as long as the sum of the values stored in the table~$T$ does not exceed $|V|$, the total number of agents.

 In the first phase (lines \ref{Line:phase1-start}--\ref{Line:phase1-end}), we construct a network as described in \cref{Cons:Graph} and check whether it admits a \emph{perfect} flow; see \cref{def:flow-perfect-asignment}.
 If this is the case, we derive the corresponding assignment and return it (line \ref{Line:assignment}).
 Otherwise, we update the table entries of some posts by finding an obstruction (see \cref{Def:Obs}) in the second phase~(lines \ref{Line:Alg_Obs_Start}--\ref{Line:Alg_Obs_End}).
 In the third phase (lines~\ref{Line:ForeachToInvalid}--\ref{Line:Iteration-End}), we update the table~$T$ and invalidate some tuples according to the obstruction we found in the second phase.

 The correctness is based on \cref{lem:d=T[j]+invalid,lem:list-non-empty,lem:obstruction,lem:CP-assignment,lem:return-correct}.
 Before we show them, we introduce the concept of obstructions, which are witnesses for the absence of a perfect flow.
 
\begin{definition}[Obstruction]\label{Def:Obs}
A pair~$(A', V')\subseteq \NETA\times \NETV$ is called an \myemph{obstruction} for a network $N=(G,\capa)$ with $G=(\NETA\cup \NETV \cup \{s,t\}, E)$ if the following holds.
\begin{compactenum}[(i)]
  \item\label{obstruction:cardV'} $\emptyset \neq V'\subseteq \NETV$;
  \item\label{obstruction:A'} $A'=\{\hata \in \NETA \mid \exists \hatv \in V'\text{ with } (\hata, \hatv) \in E(G)\}$;
 \item\label{obstruction:bad}
For each~$\hata\in A'$ it holds that
  $\capa(s, \hata) < |\{\hatv \in V' \mid (\hata, \hatv)\in E\}|$.
  \item\label{obstruction:bad2}
  $\sum_{\hata\in A'}\capa(s, \hata) < |V'|$.  
\end{compactenum}

  Slightly abusing the terminology, $A'$ can be seen as a \emph{minimal} set of posts~$a$ with congestions that are not enough to accommodate all agents from~$V'$. 
\end{definition}

Throughout the remainder of the section, by an iteration, we mean the execution of lines~\ref{Line:Iteration-Start}--\ref{Line:assignment} if a CP assignment is found, or lines~\ref{Line:Iteration-Start}--\ref{Line:Iteration-End} otherwise.
In order to refer to the specific table in an iteration, we enumerate the tables.
For each iteration~$z$, $z\ge 1$, we use \myemph{$T_z$} to denote the table at the beginning of iteration~$z$ (i.e., at line~\eqref{Line:Iteration-Start}).

The next lemma ensures that increasing the congestions for the posts correspond to the most preferred valid tuples.
\begin{lemma}\label{lem:d=T[j]+invalid}
  \begin{enumerate}[(i)]
    \item\label{lem:stillvalid} At the beginning of each iteration~$z$, if $(a,d)$ is valid,
    then $T_{z}[a] \le d$. 
    \item\label{lem:best-valid-tuple}  At the beginning of each iteration~$z$, if $(a, d)$ is the most preferred valid tuple of some agent,
    then $T_z[a]=d$.
    \item\label{lem:invalid} If a tuple~$(a,d)$ is set invalid during an iteration~$z$, then
    all tuples $(a,d')$ with $d'\in [d-1]$ are invalid and 
    and $T_{z+1}[a]=d+1$.
  \end{enumerate}
\end{lemma}

\begin{proof}
  Statement~\eqref{lem:stillvalid} clearly holds for the first iteration due to line~\ref{Line:InitTj}.
  Now, consider any further iteration~$z$.
  By lines~\ref{Line:SetToInvalid}--\ref{Line:SetToIncrement}, right before
  a table entry~$T[a]$ is increased by one, the tuple~$(a, T[a])$ is set to invalid.
  This means that $T_z[a]-1$ many tuples containing $a$ were set invalid before the $z^{\text{th}}$ iteration.
  The corresponding congestions in the invalid tuples are $1, 2, \ldots, T_z[a]-1$.
  Hence, $T_z[a] \le d$, confirming the first statement.

  By statement~\eqref{lem:stillvalid}, in order to show the second statement, it suffices to show that if $(a,d)$ is the most preferred \emph{valid} tuple of some agent, then $T_z[a] \ge d$.
  This is clearly the case for $z=1$ due to line~\ref{Line:InitTj} and the preferences are negative towards congestion.
  Suppose, for the sake of contradiction, $T_{z'}[a] \ge d$ holds for all iterations~$z' \in [z-1]$, but at the beginning of iteration~$z$ we have $T_{z}[a] < d$.
  Since the table entries never decrease, we have that $d \le T_{z'}[a] \le T_z[a] < d$, a contradiction.
  This proves statement~\eqref{lem:best-valid-tuple}. 

  For statement~\eqref{lem:invalid}, let $(a,d)$ be a tuple that is set invalid during iteration~$z$.
  Then, by our algorithm in lines~\ref{Line:Iteration-Start}, $(a,d)$ must be the most preferred valid tuple of some agent at the beginning of iteration~$z$.
  By statement~\eqref{lem:best-valid-tuple}, we have that $T_z[a] = d$.
  Since we set a tuple invalid whenever we increase the table entry,
  in iteration~$z$, all tuples $(a, d')$ with $d'\in [T_z[a]-1] = [d-1]$ are already invalid, showing the first half of the statement.
  The second half follows analogously.
\end{proof}

The next lemma guarantees the correctness of the while-loop condition.
\begin{lemma}\label{lem:list-non-empty}
  For each iteration~$z$, if $\sum_{a\in A}T_z[a] \le |V|$, then each agent has at least one valid tuple in his preference list.
\end{lemma}

\begin{proof}
  Let $z$ and $T_z$ be as described in the if-condition.
  Let $n$ denote the number of agents.
  Then, each agent has exactly $n$ valid tuples in his preference list at the beginning of the algorithm.
  Suppose, for the sake of contradiction, that some agent~$v$'s preference list contains only invalid tuples.
  By \cref{lem:d=T[j]+invalid}\eqref{lem:invalid}, this means that for each acceptable post $a\in A$ and each $d\in [\MaxCong{v}{a}]$
  the tuple $(a,d)$ was set invalid at some iteration.
  By the same lemma, we obtain that $T_{z}[a]\ge \MaxCong{v}{a}+1$.
  This implies that $\sum_{a\in A}T_z[a] \ge  \big(\sum_{a\in A}\MaxCong{v}{a}\big)+1 > n$, a contradiction to the if-condition.
\end{proof}

The next lemma guarantees that the second phase is doable.

\begin{lemma}\label{lem:obstruction}
 Each $(A', V')$ computed at the end of the repeat-loop in lines~\ref{Line:Alg_Obs_Start}--\ref{Line:Alg_Obs_End} is an obstruction.
\end{lemma}

\begin{proof}
  Let $(A', V')$ be the pair computed in lines~\ref{Line:Alg_Obs_Start}--\ref{Line:Alg_Obs_End} in some iteration~$z$.
  Let $(G, \capa)$ be the network with $G=(\NETA\cup \NETV\cup \{s,t\}, E)$ and $f$ the maximum flow computed in iteration~$z$.
  We aim to show that $(A', V')$ satisfies the properties defined in~\cref{Def:Obs}.

  By the algorithm, $f$ fails to have value~$|V|$, i.e., $\sum_{\hatv\in \NETV}f(\hatv, t) < |V|$.
  Hence, there must be a vertex~$\specialv\in \NETV$ with $f(\specialv,t)=0$.
  Let $\specialv$ be such a vertex that is added to $V'$ in line~\ref{Line:V'-init}.
  Then, property~\eqref{obstruction:cardV'} is clear since $\specialv\in V'$ (line~\ref{Line:V'-init}) and we only add vertices from $\NETV$ to $V'$; see line~\ref{Line:V'}.

  Property~\eqref{obstruction:A'} is also clear due to line~\ref{Line:A'}.
  
  Let us consider property~\eqref{obstruction:bad}.
  Clearly, for every vertex~$\hat{a}\in A'$ with $(\hat{a}, \specialv)\in E(G)$
  we must have that $f(s, \hata) = \capa(s, \hata)$ as otherwise we can increase the flow by one by setting $f(s,\hata)=f(s,\hata)+1$ and $f(\hata, \specialv)=f(\specialv, t)=1$.
  By line~\ref{Line:V'}, every out-neighbor~$\hat{v}$ with $f(\hata, \hatv)=1$ is added to $V'$.
  Together with $\specialv$, we obtain that $\capa(s, \hata) < |\{\hatv\in V' \mid (\hata, \hatv)\in E\}|$, as desired.

   Now, consider an arbitrary vertex~$\hata\in A'$ with $(\hata, \specialv)\notin E(G)$.
   Suppose, towards a contradiction, that $\hata$ does not satisfy the last property, meaning that $\capa(s,\hata) \ge |\{\hat{v'}\in V' \mid (\hata, \hat{v'}) \in  E(G)\}|$.
   We aim to show that there is an ``augmenting'' path from $\hata$ to $\specialv$, with arcs having flow values alternating between zero and one, which is a witness for the flow to be not maximum.
   
   Let us review the repeat-loop in lines~\ref{Line:Alg_Obs_While}--\ref{Line:Alg_Obs_End}.
   Observe that in each round of this loop, we aim at finding a vertex~$\hat{a}$ not already from $A'$ that has an in-neighbor from some vertex from~$V'$; $V'$ is initialized with $V'=\{\specialv\}$.
   This means that we can find a vertex in~$\hat{v}_{x}\in V'\setminus \{\specialv\}$ due to which we add~$\hat{a}$.
   Further, for each vertex~$\hat{v'}$ in $V'\setminus \{\specialv\}$, we can also find a vertex~$\hat{a'}$ in a previous round such that $f(\hat{a'}, \hat{v'})=1$.
   Let $\hat{a}_x$ be the vertex from $A'$ due to which we add $\hat{v}_x$, i.e., $f(\hat{a}_x, \hat{v}_x)=1$.
   Since each vertex has only one out-arc with capacity one, due to the conservation constraint of the flow~$f$,  we infer that $f(\hat{a}, \hat{v}_{x})=0$. 
   Repeating the above reasoning, there must be a vertex~$\hatv_{x-1}$ from $V'\setminus \{\hat{v}_x\}$ due to which we add $\hat{a}_x$.
   Then, either $\hat{v}_{x-1}=\specialv$ or $\hat{v}_{x-1}\neq \specialv$.
   In the former case, we infer that $(\hata, \hatv_x, \hata_x, \specialv)$ is an augmenting path, so flipping the arc flows as follows would increase the value of the flow:
   \begin{align*}
     f(s, \hat{a})=f(s, \hat{a})+1, 
     f(\hat{a}, \hatv_{x}) = 1, \\
     f(\hata_x, \hatv_x) = 0, f(\hata_x, \specialv) = f(\specialv, t) = 1,
   \end{align*}
   a contradiction.
   In the latter case, since $V'$ is finite and no vertex from $\NETV\supseteq V'$ can have more than one positive flow, by repeating the above reasoning, we must end up with an arc to $\specialv$ with zero flow; recall that $\specialv\in V'$.
   Then, we again obtain an augmenting path~$P=(\hata, \hatv_x, \hata_x, \ldots, \hata_0, \hatv_0=\specialv)$.
   Analogously, we can increase the flow by flipping the arc flows along this path. A contradiction.     

   It remains to show property~\eqref{obstruction:bad2}.
   This is clear since otherwise for each vertex~$\hatv\in V'$ we can find a vertex~$\hata\in A'$ and set $f(\hata, \hatv)=f(\hatv,t)=1$.
   In particular, the starting vertex~$\specialv$ would have non-zero flow going through it, a contradiction.
\end{proof}

The next lemma ensures that increasing the table entries are safe; this is necessary for the correctness of the no-answer.
\begin{lemma}\label{lem:CP-assignment}
  Assume that $I$ admits a CP assignment with no posts being empty and let $\Pi$ denote such a CP assignment.
  Then, for each iteration~$z\ge 1$ and each post~$a\in A$,
  if we set $(\hata, T_z[a])$ invalid in line~\ref{Line:SetToInvalid},
  then $|\Pi(a)|\ge T_z[a]+1$; otherwise $|\Pi(a)|\ge T_z[a]$.
\end{lemma}

\begin{proof}
  Let us consider the first iteration, clearly, the statement holds for posts~$a$ whose tuples are not set invalid since $\Pi$ does not have empty posts.
  By our algorithm, the invalidated tuples contain posts that are in the obstruction $(A', V')$ computed in lines~\ref{Line:Alg_Obs_Start}--\ref{Line:Alg_Obs_End}.
  Let $N=(G, \capa)$ denote the network and $f$ the maximum flow of $N$ computed in the first phase.
  For the sake of reasoning, let $P=\{v\in V \mid \hatv\in V'\}$ be the set of agents that correspond to the vertices in $V'$ and $Q=\{a\in A \mid \hata \in A'\}$. 
  
  Suppose, for the sake of contradiction, that there exists a vertex~$a\in A'$ with $|\Pi(a)| \le T_1[a]=1$.
  By \cref{Def:Obs}\eqref{obstruction:bad}, more than $\capa(s, \hata)=1$ vertices from $V'$ are out-neighbors of $\hata$ in~$G$.
  By definition, at least two agents from $P$ consider $(a,1)$ as one of the most preferred tuples.
  
  Since $|\Pi(a)|=1$, at least one agent from $P$ is not assigned to $a$ but considers $(a, 1)$ as one of the most preferred tuples.
  Let $v_0\in P$ be such an agent.
  Then, he must be assigned to some other post~$a_0$ such that $(a_0, |\Pi(a_0)|)$ is one of the most preferred tuples for~$v_0$ as well.
  This means that $|\Pi(a_0)|=1$.
  By \cref{Cons:Graph}, we infer that $(\hata_0, \hatv_0)\in E(G)$, and by \cref{Def:Obs}\eqref{obstruction:A'} that $\hata_0\in A'$.

  Analogously, by \cref{Def:Obs}\eqref{obstruction:bad}, more than $\capa(s, \hata)=1$ vertices from $V'$ are out-neighbors of $\hata_0$ in~$G$,
  and we can find another agent~$v_1\in P$ that is not assigned to $a_0$ but considers $(a_0, 1)$ one of the most preferred tuples.
  Again, this agent~$v_1$ will be assigned to some post~$a_1$ with $(a_1, 1)$ being one of the most preferred tuples of $v_1$.
  By \cref{Cons:Graph}, we infer that $(\hata_1, \hatv_1)\in E(G)$, and by \cref{Def:Obs}\eqref{obstruction:A'} that $\hata_1\in A'$.
  Repeating the above reasoning, we will be able to find a distinct vertex~$\hata_i\in A'$ for each vertex~$\hatv_i\in V'$.
  That is, $|A'| \ge |V'|$, a contradiction to \cref{Def:Obs}\eqref{obstruction:bad2}
  since $|A'| = \sum_{\hata\in A'}\capa(s, \hata)$ in this case.

  Now, let us consider any further iteration.
  By our algorithm, the table entries never decrease.
  Hence, if the statement were incorrect, there must be an iteration~$z \ge 2$ where the statement holds in iteration~$z-1$ but not in iteration~$z$.
  
  Suppose, for the sake of contradiction, that the statement is incorrect; let $z$ be the first iteration index and $a$ one of the first posts such that $|\Pi(a')|\ge T_{z-1}[a']$ holds for all $a'\in A$ but $|\Pi(a)| < T_{z}[a]$.
  Since in each iteration, the table entry of each post is increased by at most one but never decrements, it follows that $|\Pi(a)| = T_{z-1}[a] = T_{z}[a]-1$. %
  By the for-loop in lines~\ref{Line:ForeachToInvalid}--\ref{Line:SetToIncrement},
  $(a, T_z[a])$ is set invalid in iteration~$z-1$ due to a pair~$(A', V')$ computed in lines~\ref{Line:Alg_Obs_Start}--\ref{Line:Alg_Obs_End} with $\hata \in A'$.
  Then, let $(G, \capa)$ denote the network constructed in iteration~$z-1$.
  By \cref{lem:obstruction}, $(A', V')$ is an obstruction for $(G, \capa)$  with $\specialv\in V'$ being the starting vertex with $f(\specialv, t)=0$.
  By \cref{Def:Obs}\eqref{obstruction:bad}, more than $\capa(s, \hata) = T_{z-1}[a]=|\Pi(a)|$ vertices from $V'$ exist that have $a$ as in-neighbors.
  Slightly abusing the interpretation of $V'$, by \cref{Cons:Graph}, at least $|\Pi(a)|+1$ agents from $V'$ consider $(a, d)$ as one of the most preferred valid tuple in iteration~$z-1$.
  By \cref{lem:d=T[j]+invalid}\eqref{lem:best-valid-tuple}, $d=T_{z-1}[a]=|\Pi(a)|$.
  Hence, at least one of such agents is not assigned to~$a$ by~$\Pi$.

  Let $\hat{v}\in V'$ be a vertex such that the corresponding agent~$v$ considers $(a, |\Pi(a)|)$ as one of the most preferred valid tuple but is assigned to some other post~$a'\neq a$.
  Then, $(\hat{a}, \hatv)\in E(G)$. %
  To prevent $v$ from being envious (recall that no post is empty), we must have that
  $(a', |\Pi(a')|)\succeq_{v} (a, |\Pi(a)|)$.
  Since $(a, |\Pi(a)|)$ is one of the most preferred valid tuple for~$v$ and $\Pi$ is CP,
  $(a', |\Pi(a')|)$ must also be one of the most preferred valid tuple for~$v$.
  By \cref{lem:d=T[j]+invalid}\eqref{lem:best-valid-tuple}, it follows that $T_{z-1}[a']=|\Pi(a')|$.  
  Further, since $(a, |\Pi(a)|) \sim_{v} (a', |\Pi(a')|)$, by the construction of the network, we have that $(\hat{a'}, \hatv)\in E(G)$.  
  By \cref{Def:Obs}\eqref{obstruction:A'} and since $\hat{v} \in V'$, we have that $\hat{a'}\in A'$. 
  By \cref{Def:Obs}\eqref{obstruction:bad}, we infer that more than $\capa(a')=T_{z-1}[a']=|\Pi(a')|$ vertices from $V'$ have an in-arc from $\hat{a'}$.
  By the construction of flow network,  more than $\capa(a')=T_{z-1}[a']=|\Pi(a')|=|\Pi(a')|$ agents consider $(a', |\Pi(a')|)$ as one of the most preferred valid tuple.

  By an analogous reasoning as above, we can again find another vertex~$\hat{v'}\in V'$
  such that $v'$ considers $(a', |\Pi(a')|)$ as one of the most preferred valid tuple but is assigned to some other post~$a''\neq a'$ with $\hat{a''}\in A'$ and $T_{z-1}[a'']=|\Pi(a'')|$.
  By continuing this reasoning, we obtain that every vertex~$\hat{\alpha}\in A'$ has $T_{z-1}[\alpha]=|\Pi(\alpha)|$.
  By \cref{Def:Obs}\eqref{obstruction:bad2}, we have that $|V'| > \sum_{\hat{\alpha}\in A'}\capa(s, \hat{\alpha}) = \sum_{\hat{\alpha}\in A'}T_{z-1}[\alpha] = \sum_{\hat{\alpha}\in A'}|\Pi(\alpha)|$.
  So there must be a vertex~$\hat{\mu}\in V'$ such that $\mu$ is assigned to a post~$b$ with $\hat{b}\notin A'$.
  Since no agent's preference list contains only invalid tuples (see \cref{lem:list-non-empty}),
  there must exist a vertex~$\hat{\alpha}\in A'$ with $(\hat{\alpha}, \hat{\mu})\in E(G)$, meaning that $(\alpha, |\Pi(\alpha)|)$ is one of the most preferred valid tuple of $\mu$.
  By definition, all valid tuples that are more preferred to $(\alpha, |\Pi(\alpha)|)$ by $\mu$ are set invalid.
  Hence, $(\alpha, |\Pi(\alpha)|) \succeq_{\mu} (b, |\Pi(b)|)$.
  Since $\hat{b}\notin A'$, by \cref{Cons:Graph} and \cref{Def:Obs}\eqref{obstruction:A'}, we have that $(\hat{b}, \hat{\mu})\notin E(G)$, meaning that $(b, |\Pi(b)|)$ is \emph{not} one of the most preferred valid tuple of $\mu$. 
  This means that $(\alpha, |\Pi(\alpha)|) \succ_{\mu} (b, |\Pi(b)|)$, a contradiction to $\Pi$ being competitive. 
\end{proof}

The next lemma ensures that the returned assignment is competitive. 
\begin{lemma}\label{lem:return-correct}
  If $\Pi$ is an assignment returned in line~\ref{Line:assignment} during iteration~$z$,
  then it is competitive for $I$ with all posts being non-empty.
\end{lemma}

\begin{proof}
  Let $\Pi$ and $z$ be as defined.
  Further, let $f$ be the perfect flow based on which $\Pi$ is computed; see line \ref{Line:assignment}.
  By the definition of perfectness (see \cref{def:flow-perfect-asignment}), the value of $f$ equals the number~$|V|$ of agents.
  This means that $\sum_{a\in A}|\Pi(a)| = |V|$.
  By the capacity constraints, we obtain that
  \begin{align*}
    |V|=\sum_{a\in A}|\Pi(a)| \le \sum_{a\in A}T_z[a] \le |V|,
  \end{align*}
  the last inequality holds due to the while-loop-condition in line~\ref{Line:SumTj<=n}.
  Hence, for each post~$a\in A$ we must have that $|\Pi(a)|=T_z[a]$ since $|\Pi(a)| \le T_z[a]$ holds by the capacity constraints.

  This immediately implies that $\Pi(a)\neq \emptyset$ since $T_z[a]\ge 1$.
  Hence, in order to show that $\Pi$ is competitive, it suffices to show that for each agent~$v$ that is assigned to a post~$a$ and each acceptable post~$a'$ with $a'\neq a$
  we have that 
  $(a, |\Pi(a)|) \succeq_v (a', |\Pi(a')|)$.
  Suppose, for the sake of contradiction, that $(a', |\Pi(a')|) \succ_v (a, |\Pi(a)|)$.
  Then, by our algorithm, this implies that $(a', |\Pi(a')|)$ is an invalid tuple and was set invalid during some previous iteration.
  This further implies that $T_z[a']\ge |\Pi(a')|+1$, a contradiction to our previous reasoning.
\end{proof}

The above five lemmas yield the correctness of our algorithm.

\begin{theorem}\label{thm:noEmpty}
  \cref{Alg:CP_Ties_NoEmpty} correctly decides whether a given instance~$I$ admits a CP assignment with no post being empty in $O(m\cdot n^2)$ time, where $m$ and $n$ denote the number of posts and agents, respectively.
\end{theorem}

\begin{proof}
  Let $I$ denote an instance with $I=(A, V, (\succeq_v)_{v\in V})$, $m=|A|$, and $n=|V|$.
  Clearly, if our algorithm returns an assignment~$\Pi$, then by \cref{lem:return-correct}, 
  it is competitive and every post is assigned at least one agent.
  
  Hence, to show the correctness, we need to show that whenever our algorithm returns no, $I$ does not admit a CP assignment where every post is non-empty.
  Towards a contradiction, suppose that $I$ admits a CP assignment with no post being empty, say $\Pi$.
  Since our algorithm returns no, in the second last iteration~$z$, we have $\sum_{a\in A}T_{z}[a] \ge |V|$, but after the update of some table entries the sum exceeds $|V|$.
  Let $T_{z+1}$ denote the table entries at the end of iteration~$z$.
  By assumption, we have that $\sum_{a\in A}T_{z+1}[a] > |V|$ and that $\sum_{a\in A}T_z[a]\le |V|$.
  
  This means that we computed an obstruction~$(A', V')$ in iteration~$z$.
  Then, by \cref{lem:CP-assignment}, we have that $|\Pi(a)|\ge T_z[a]+1$ for all $\hata \in A'$.
  Again, by \cref{lem:CP-assignment}, $|\Pi(a')|\ge T_z[a']$ holds for all $a' \in A$ with $\hat{a'}\notin A'$.
  By lines~\ref{Line:ForeachToInvalid}--\ref{Line:SetToIncrement}, we obtain that $|\Pi(a)|\ge T_{z+1}[a]$ holds for all $a\in A$.
  Then, $\sum_{a\in A}|\Pi(a)|\ge \sum_{a \in A}T_{z+1}[a] > |V|$, a contradiction to $\Pi$ being a CP assignment.

  It remains to analyze the running time.
  Initializing the table~$T$ and validating all tuples needs $O(m)$ time and $O(n\cdot m)$ time, respectively.
  The while-loop (lines~\ref{Line:SumTj<=n}--\ref{Line:Iteration-End}) runs at most $n$ times since no table entries are ever decreased and in each iteration at least one table entry is increased by one.
  For each iteration, we first construct a network~$N=(G,\capa)$ based on $(I, T)$; see \cref{Cons:Graph}.
  The directed graph~$G$ has $O(n+m)$ vertices and $O(m\cdot n)$ arcs, and the capacity function~$\capa$ can be computed in $O(m\cdot n)$ time.
  Hence, constructing the network needs $O(m\cdot n)$ time.

  Afterwards, there are three phases.
  The first phase (lines~\ref{Line:phase1-start}--\ref{Line:phase1-end}) finds a maximum flow for $N$ and check whether the value is $|V|$.
  Computing a maximum flow can be done in $O(m\cdot n)$ time and comparing the value in $O(n+m)$ time.
  Hence, the first phase needs $O(m\cdot n)$ time.

  The second phase (lines~\ref{Line:Alg_Obs_Start}--\ref{Line:Alg_Obs_End}) finds an obstruction~$(A', V')$ by first finding a vertex~$\specialv$ with $f(\specialv, t)=0$.
  This can be done in $O(1)$ time if we store such information when we compare the value of the flow with $|V|$ in the first phase.
  Hence, the initialization of $V'$ and $A'$ needs $O(1)$ time.
  Then, the algorithm goes to the repeat-loop in lines~\ref{Line:Alg_Obs_While}--\ref{Line:Alg_Obs_End}.
  To analyze the running time for this loop, we observe that there are $O(m\cdot n)$ arcs between $A'$ and $V'$ and each arc only needs to be checked at most once during the whole loop (line \ref{Line:A'}).
  Adding new vertices to $V'$ can be done in $O(m\cdot n)$ time as well since for each newly added alternative~$\hata$ there are at most $n$ vertices~$\hatv$ from $\NETV$ with positive flow from~$\hata$ to~$\hatv$.
  Hence, the repeat loop needs $O(m\cdot n)$ time.

  It is straightforward that the last phase (lines~\ref{Line:ForeachToInvalid}--\ref{Line:Iteration-End}) runs in $O(m)$ time.
  Overall, the algorithm runs in $O(m\cdot n^2)$ time. 
\end{proof}

\subsection{Correctness of Algorithm~\ref{Alg:CP_Ties}}\label{subsec:alg1}

We have everything ready to show the correctness of \cref{Alg:CP_Ties}.

\begin{theorem}\label{thm:main}
  Algorithm~\ref{Alg:CP_Ties} correctly decides whether an instance has a CP assignment in $O(m^2\cdot (n+m)^2)$ time, where $m$ and $n$ denote the number of posts and agents, respectively.
\end{theorem}

\begin{proof}
  The correctness of the algorithm follows directly from \cref{lem:I<=>Ik} and \cref{thm:noEmpty}. 
  
  It remains to analyze the running time.
  The main body of the algorithm is a for-loop (lines~\ref{alg1-for}) and runs at most $m$ iterations.
  In each iteration, the algorithm constructs a new instance~$I_k$ according to \cref{Cons:instance}  and  \cref{Alg:CP_Ties_NoEmpty}.
  Note that $I_k$ has $O(n+m)$ agents and $O(m)$ posts, and it can be constructed in $O((n+m)^2)$ time since each agent has $O(n+m)$ tuples in his preference list.
  Afterwards, the algorithm calls \cref{Alg:CP_Ties_NoEmpty} on instance~$I_k$ which can be done in $O(m\cdot (n+m)^2)$ time by \cref{thm:noEmpty}.
  Summarizing, the algorithm runs in $O(m^2\cdot (n+m)^2)$ time. 
\end{proof}

\section{NP-Hardness for EF Assignments}\label{sec:EF}

In this section, we focus on envy-freeness and show that is is NP-hard to obtain.
We reduce from the following NP-hard problem.
 \decprob{\ExactCover}
 {A finite set~$U = \{u_1, \ldots , u_{3n}\}$ of $3n$ elements , subsets~$S_1, \ldots, S_m \subseteq U$ with $|S_j| = 3$ for each~$j \in [m]$.}
 {Does there exist an \myemph{exact cover $J \subseteq [m]$} for $U$, i.e., $\bigcup_{j \in J} S_j = U$ and each element occurs in exactly one set in the cover?}

 This problem remains NP-hard even if each element appears in exactly three subsets. 
 
 \begin{theorem}\label{thm:EF-NP-hard}
   It is NP-complete to decide whether an instance of \Congested admits an EF assignment.
 \end{theorem}

 \begin{proof}
   NP-containment is clear since one can check in polynomial time whether a given assignment is envy-free.
   Now, we focus on NP-hardness and reduce from the NP-complete \ExactCover problem.
   Let $I=(U, \mathcal{S})$ denote an instance of \ExactCover with $U=\{u_1,\ldots,u_{3n}\}$and $\mathcal{S}=(S_j)_{j\in [m]}$ such that every element in $U$ appears in exactly three members of $\mathcal{S}$.
   This means that $m=3n\ge 3$.
   
   We create an instance of \Congested as follows.
   \begin{compactitem}[--]
     \item For each member~$S_j\in \mathcal{S}$, create a \myemph{set-post}~$a_j$;
     \item For each element~$u_i\in U$, create an \myemph{element-agent}~$v_i$;
     \item Create two dummy posts~$b_1$ and $b_2$ and $4m$ dummy agents~$p_1, p_2, \ldots, p_{2m}, q_1, q_2, \ldots, q_{2m}$.
   \end{compactitem}
   Let $A=\{a_j\mid j\in [m]\}\cup \{b_1,b_2\}$ and $V=\{v_i\mid i\in [3n]\}\cup \{p_j\mid j \in [2m]\}$.
   
 \paragraph{Preferences.} We describe the preferences of the agents.
   \allowdisplaybreaks
   \begin{compactitem}[--]
     \item The preferences of agent~$v_i$ is defined as follows, where $S_j, S_k, S_t$ denote the three members in $\mathcal{S}$ that contain~$u_i$:
   \begin{align*}
     v_i \colon  &(a_j,1) \sim (a_k,1) \sim (a_t, 1) \succ (a_j,2) \sim (a_k,2) \sim (a_t, 2) \\
                 & \succ (a_j,3) \sim (a_k,3) \sim (a_t, 3) \succ (b_2, 1)\succ (b_2, 2)\\
     & \succ \ldots \succ (b_2,3n+4m-8).
   \end{align*}
   In other words, each agent~$v_i$ consider the three posts which correspond to the sets that contain $u_i$ most acceptable, followed by $b_2$.
   He does not consider any other post acceptable.

  \item Each dummy agent~$p_j$, $j\in [2m]$, has the same preferences:
   \begin{align*}
     p_j\colon & (a_1,1)\sim (a_2, 1) \sim \ldots  \sim (a_m, 1) \succ \\
               & (a_1,2)\sim (a_2, 2) \sim \ldots  \sim (a_m, 2)\succ \\
               & (b_1,1)\succ (b_1,2)\succ \ldots \succ (b_1, 2m+3n).
   \end{align*}
   Briefly put, each dummy agent~$p_j$ always want to go to a set-post with congestion one or two.

   \item Each dummy agent~$q_j$, $j\in [2m]$, only consider $b_1$ and $b_2$ acceptable, but prefers $b_2$ over $b_1$:
   \begin{align*}
     q_j\colon & (b_2, 1) \succ (b_2, 2) \succ \ldots (b_2, 2m)\succ\\
      & (b_1, 1) \succ (b_1, 2) \succ \ldots \succ (b_2, 2m+3n).
   \end{align*}
 \end{compactitem}
 
   The maximum congestions of the agents are depicted in the following table, where $v_i$ is an element-agent with $u_i$ appearing in $S_1, S_2, S_m$:
   
   \noindent
   \begin{tabular}{@{}|l@{\;}|lll@{\;\;}l@{\;\;}l@{\;\;}|ll@{\;}|@{}}\hline
    & $a_1$ & $a_2$ & $a_3$ & $\ldots$ & $a_m$ & $b_1$ & $b_2$ \\\hline
    $v_i$ & $3$ & $3$ & $0$ & $\ldots$ & $3$ & $0$ & $3n+4m-9$\\\hline
    $p_z$ & $2$ & $2$ & $2$ & $\ldots$  & $2$  & $3n+2m$ & $0$\\\hline
    $q_z$ & $0$ & $0$ & $0$ & $\ldots$  & $0$  & $3n+2m$ & $2m$\\\hline
  \end{tabular}
  \smallskip

  \paragraph{Correctness.}  
  Clearly, the construction can be done in linear time.
   Let $I'$ denote the constructed instance.
   It remains to show the correctness, i.e., $I$ has an exact cover if and only if $I'$ admits an EF assignment.

   For the ``only if'' part, let $J\subseteq [m]$ denote an exact cover for $I$.
   Then, we claim that the following assignment~$\Pi$ is envy-free:
   \begin{compactitem}[--]
     \item For each~$j\in J$, let $\Pi(a_j) = \{v_i \mid u_i \in S_j\}$.
     \item For each~$j\in [m]\setminus J$, let $\Pi(a_j) = \emptyset$.
     \item Let $\Pi(b_1)=\{p_j \mid j\in [2m]\}$ and $\Pi(b_2)=\{q_j\mid j\in [2m]\}$.
   \end{compactitem}
   Since each set-post contains either zero or three agents, no dummy agent envies any element-agent.
   The dummy agents also do not envy each other due to their preferences.
   Similarly, no two element-agents envy each other and no element-agent envies any dummy agent since he does not like $b_1$ or $b_2$ more.

   For ``if'' part, let $\Pi$ be an envy-free assignment for the constructed instance.
   We aim at showing that the set-posts that are assigned element-agents constitute an exact cover.
   To this end, let $J=\{j\mid \exists v_i \text{ with } v_i\in \Pi(a_j)\}$.
   We first show claim that helps to confirm that $J$ is an exact cover.

   \begin{claim}\label{claim:set-0/3}
     For each set-post~$a_j$ it holds that $|\Pi(a_j)|\in \{0,3\}$.
   \end{claim}
   \begin{proof}        \renewcommand{\qedsymbol}{%
       $\diamond$}
     Since there are $2m$ dummy agents~$\{p_1, p_2, \ldots, p_{2m}\}$, but there are only $m$ set-posts,
     at least one dummy agent, say $p_{z}$, is not assigned to a set-post alone.
     Hence, for every set-post~$a_j$, it holds that $|\Pi(a_j)| \neq 1$ since otherwise $p_{z}$ will envy the agent that is assigned to $a_j$.
     Since the maximum congestion for every set-post is $3$, we further infer that $\Pi \in \{0,2,3\}$ holds for every set-post~$a_j$.
     In particular, this implies that no dummy agent~$p_{z}$ is assigned to a set-post alone.
         
     Towards a contradiction, suppose that there exists a set-post~$a_j$ with $|\Pi(a_j)|\notin \{0,3\}$.
     This implies that $|\Pi(a_j)| = 2$ since no agent will the post with $4$ or larger congestion.
     Then, no dummy agent~$p_{z}$ can be assigned to a post that is not a set-post since otherwise he will the two agents that are assigned to $a_j$.
     Since there are exactly $2m$ dummy agents, this means that every set-post~$a_{x}$, $x\in [m]$, must have $|\Pi(x)|=2$.
     Then, no other agent can be assigned to the set-post anymore due to the congestion constraints.
     However, all element-agents will envy all dummy agents, a contradiction.
     This concludes the proof.
   \end{proof}

   By \cref{claim:set-0/3}, we know that each set-post is assigned either zero or three agents.
   Next, we show that every element-agent is assigned to an acceptable set-post.
   \begin{claim}\label{claim:setcover}
     For each element-agent~$v_i$ it holds that $\Pi(v_i) \in \{a_j \mid j\in [m]\}$.
   \end{claim}
   \begin{proof}
     \renewcommand{\qedsymbol}{$\diamond$}
     Suppose that this is not true, and let $v_i$ denote an element-agent that is assigned to~$b_2$; note that $v_i$ does not find $b_1$ acceptable.
     Then, since there are $2m$ dummy agents~$q_z$ each with congestion~$2m$ for $b_2$, at least one of them is \emph{not} assigned to~$b_2$.
     This agent envies $v_i$, a contradiction.
   \end{proof}

   \cref{claim:setcover} implies that $J$ is a set cover, while \cref{claim:set-0/3} implies that $|J|\le n$.
   Altogether we conclude that $J$ is an exact cover.
 \end{proof}

 \clearpage
\bibliography{bib}

\end{document}